\documentclass[12pt]{article}
\usepackage{latexsym,epsfig}
\usepackage{graphicx}
\usepackage{pifont}
\usepackage{verbatim}
\usepackage{caption}
\usepackage{subcaption}
\usepackage{color}
\usepackage{soul}
\usepackage[utf8]{inputenc}
\usepackage[T1]{fontenc} 
\usepackage{lmodern} 
\usepackage{cite}
\def\ed{\end{document}}

\usepackage{color}
\def\sk{\smallskip}

\def\bg{\bigskip}

\def\beq{\begin{eqnarray}}
\def\eq{\end{eqnarray}}
\def\beqn{\begin{eqnarray*}}
\def\eqn{\end{eqnarray*}}

\def\pt{\ensuremath{p_{\mathrm{T}}}} 

\begin{document}
\begin{center}
{\bf \large First results on Bilepton production based  
\sk

on LHC collision data and predictions for Run II}
\bg 

{A. A. Nepomuceno$^{1}$, B. Meirose$^{2}$ and F.L. Eccard$^{1}$}

\vskip .5 cm

{$^{1}$Departamento de Ci\^encias da Natureza, Universidade Federal Fluminense \\
Rua Recife, 28890-000 Rio das Ostras, RJ, Brazil
}%

{$^{2}$Department of Physics \\
University of Texas at Dallas  \\
800 W. Campbell Road, Richardson, Texas, USA}
\end{center}

\begin {abstract}
\noindent
The LHC potential for discovering doubly-charged vector bileptons is investigated considering the 
measurable process $p p$ $\rightarrow$ $\mu^{+}\mu^{+}\mu^{-}\mu^{-} X$. The study is performed assuming different 
bilepton and leptoquark masses. The process cross-section is calculated at leading-order using the \textsc{Calchep} 
package. Combining the calculation with the latest ATLAS experiment results at a center-of-mass energy of 7 TeV, bounds on bilepton masses based on LHC data are derived for the first time.
The results exclude bileptons masses in the range of 250 GeV to 500 GeV at 95\% C.L., depending on the leptoquark mass. Moreover, minimal LHC integrated luminosities needed for discovering 
and for setting limits on bilepton masses are obtained for 13 TeV center-of-mass energy. Simulated events are passed through a fast parametric detector simulation using the \textsc{Delphes} package.
\end {abstract}

\vfill\eject
\section{Introduction}
The first run of the LHC have discarded or disfavored several new physics scenarios, with no significant excess of events compared to the Standard Model (SM) expectations having yet been observed by any of the LHC experiments in a plethora of different final states. The most significant result, the discovery of a particle consistent with the SM Higgs boson \cite{Higgs_ATLAS, Higgs_CMS}, while outstanding, has so far only strengthened our confidence in the SM.
However, the famous SM puzzles that have motivated the pre-LHC model building era are still 
unsolved and very much alive, desperately needing guidance from experiment to be unraveled. 
\vskip .2 cm
One of the dramatic effects caused by the LHC results is that, while some beyond SM (BSM) 
searches became less appealing, others may now experience renewed interest from the particle physics community. 
A good example is the search for the so-called  bileptons.
\vskip .2 cm
Bileptons are bosons with two units of leptonic number \cite{Cuypers1996}. They couple to leptons, but not to SM quarks. Bileptons do however couple to leptoquarks, which carry both baryon and lepton numbers. Scalar  bileptons are predicted by theories with enlarged Higgs sector (such as left-right models) 
as well as by models that generate neutrino Majorana masses. 
Non-gauge vector bileptons are present in composite theories, 
while heavy gauge vector bileptons, the ones studied in this article, are present when the SM is embedded in a larger gauge group. The most important and natural class of models where vector bileptons appear are the 331 models \cite{Frampton331, Pisano331, Ng1992, 331_review}, and all calculations in this article are based on them. Our main results should however hold for other models containing vector bileptons.

\subsection{Objective, motivations and paper organization}
The objective of the present article is to study doubly-charged vector bilepton production in the 
channel $p,p$ $\rightarrow$ $\mu^{+}\mu^{+}\mu^{-}\mu^{-} X$. The main motivation is to obtain the very first 
experimental limits on bileptons based on LHC collision data. To do this, the experimental limits obtained by the 
ATLAS Collaboration
for doubly-charged Higgs production are reinterpreted for the same channel \cite{doubly_ATLAS}. 
Furthermore, the five-sigma bilepton discovery potential for LHC's run II at 13 TeV center-of-mass energy is estimated, 
which complements our previous results using the $p,p$ $\rightarrow$ $e^{\mp}e^{\mp}\mu^{\pm}\mu^{\pm} X$ 
channel \cite{Meirose&Nepomuceno}.

The paper is organized as follows. In Section \ref{sec:331} a brief review of the 331 models is given, 
with focus on the features that are most relevant to the current analysis. In  Section \ref{sec:Run1}, a brief discussion on bilepton experimental limits in light of LHC's run I results is presented. Section \ref{sec:MC} describes the Monte Carlo (MC) and detector simulation procedures. In Section \ref{sec:7TeV_limits} the 95\% C.L. experimental limits based on 7 TeV LHC collision data are presented. The doubly-charged vector bilepton discovery potential for the four-muons channel at 13 TeV is shown in  Section \ref{sec:13TeV_discovery}. Conclusions are presented in Section \ref{sec:conclusion}.

\section{331 models} 
\label{sec:331}
The 331 models are based on the gauge symmetry $SU(3)_C \otimes SU(3)_L \otimes U(1)_X$, hence their name. They can generically be classified according to how they cancel chiral anomalies. For example, 
there are anomaly-free 331 models requiring only one family of quarks and leptons, although the majority of the models  studied in the literature are three familiy models \cite{Ponce2001}. We are interested in particular versions of the three family models that predict a new neutral gauge boson $Z^\prime$ and four vector bileptons $Y^{\pm}$ and $Y^{\pm \pm}$ in the gauge sector. In addition, the fermion sector of the class of 331 models studied in the present article contains three new heavy leptoquarks: $T_1$, with  electric charge $\pm 5/3$, and $D_1$ and $D_2$ both with $\pm 4/3$ of electric charge.
\vskip .2 cm
There are several reasons why three-family 331 models are good candidates to describe Nature at the TeV scale 
and we refer the reader to Ref. \cite{Meirose&Nepomuceno} for a review. 
One of their most striking motivations is that they offer an elegant solution for SM's family replication problem. 
This is achieved through two main ingredients. The first is that the cancellation of triangle anomalies is non-trivial, 
taking place between families, which can only happen if the number of families is a multiple of three. 
The second ingredient is QCD's asymptotic freedom, that requires the number of quark generations to be less than five. 
These two conditions imply that the number of families must be exactly three. It is also interesting to mention that 331 models could provide an explanation \cite{MohCao} for the (unconfirmed) diphoton anomaly, possibly a resonance around 750 GeV, that was recently reported by the ATLAS \cite{ATLASCONF} and CMS collaborations \cite{CMSCONF}.

\vskip .2 cm
Both the exotic leptoquarks and the new gauge bosons acquire mass through spontaneous symmetry breaking (SSB) of  the $SU(3)_L \otimes U(1)_X$ gauge sector. There are distinct ways how this can be accomplished and the different possibilities of Higgs sectors define further 331 model sub-versions within the three-family 331 models. The 331 minimal model 
is a particular example of one of this sub-versions that uses minimal Higgs structure for SSB. The model continues to attract attention because it requires the bilepton and $Z^\prime$ masses to be bound in a similar way the $W$ and $Z$ masses are bound in the SM. However, either than the theoretical esthetic appeal it provides through this SM resemblance, there is no real compelling reason, neither phenomenological nor experimental, to give the minimal model any privileged treatment. Indeed, there is already circumstantial evidence that, to some degree, disfavors this particular version experimentally, even though it has not yet been fully excluded \cite{Meirose&Nepomuceno}.

\section{Bileptons before and after LHC Run I}  
\label{sec:Run1}

Even before LHC's first run, limits on vector bileptons suggested that observing those particles during run I was a rather unfavorable scenario. The reason is as follows. The two most useful mass limits for vector bileptons are: $M_Y >$ 740 GeV \cite{Tully}, 
a limit derived from experimental limits on fermion pair production at LEP and lepton-flavor charged lepton decays, and $M_Y >$ 850 GeV
\cite{Willmann, Pleitez99}, a limit established from muonium-antimuonium conversion. 	In Ref. \cite{Meirose&Nepomuceno}, we have predicted that the $5 \sigma$ discovery potential for LHC's run I at 7 TeV center-of-mass energy, using 10 $\textnormal{fb}^{-1}$ of integrated luminosity, was only around 540 GeV. Even the full 20 $\textnormal{fb}^{-1}$ of data collected in run I at 8 TeV center-of-mass energy would not have been enough to surpass those two limits. However, both limits are not general, and consequently, a discovery could not have been completely discarded. The first limit has been obtained with LEP data, and as so, it is restricted to the leptonic mixing matrix being diagonal, since in 331 models, the leptons mix by a Cabibbo-Kobayashi-Maskawa-like mixing matrix whose elements have not been measured. The second limit is even more restrictive and depends on the assumption that the bilepton couplings are flavor-diagonal. Vector bilepton 
experimental limits making use of hadronic beams are therefore obtained in this paper for the first time. 
LHC constraints on general doubly-charged scalars, not necessarily scalar bileptons, were studied in Ref. \cite{Aguila}.

\vskip .2 cm
The situation for LHC's run II is quite different. Not only will the LHC be able to probe the already  searched region by LEP in a more general way using hadronic beams, it will also probe a completely new bilepton mass region around 1 TeV. 

\section{Monte Carlo and Detector Simulations}
\label{sec:MC}
The 331 model is implemented in the \textsc{Calchep} event generator \cite{belyaev} following references \cite{hoang,dion,frampton} for bilepton trilinear gauge 
interactions, $Z^{\prime}$ couplings to fermions and bileptons interactions with leptons, respectively. 
Bileptons interactions with 331 model leptoquarks are also taken into account \cite{Meirose&Nepomuceno}. 
The implementation is validated and extensively tested for consistence and unitarity. 

\vskip .2 cm
\textsc{Calchep} is used to calculate cross-sections and to produce events for several bilepton mass points for bilepton pair production. 
The generated events are processed by \textsc{Pythia} 8 \cite{Pythia, Pythia8} for hadronization and decays. A fast detector simulation is performed using \textsc{Delphes} \cite{delphes}. 
The \textsc{Delphes} package is provided with different configurations to simulate the ATLAS or CMS responses. In this work, the Snowmass Combined LHC Detector configuration is used, 
which is a general detector simulation combining ATLAS and CMS features \cite{snowmass}.
The leptoquark masses are assumed to be between 100 GeV and 800 GeV, and the $Z^{\prime}$
mass to be 3 TeV. This $Z^{\prime}$ mass  value is chosen so that it is slightly above the current experimental limits \cite{zprime}.
For bileptons, the mass range considered is 200 GeV to 1000 GeV in steps of 100 GeV. The CTEQ6L1 \cite{CTEQ6} parton distribution function (PDF) set is used in the calculations.

\section{LHC Run I: 7 TeV bilepton experimental limits}
\label{sec:7TeV_limits}

\begin{figure}
\centering
\begin{subfigure}{.5\textwidth}
  \centering
  \includegraphics[width=1.0\linewidth]{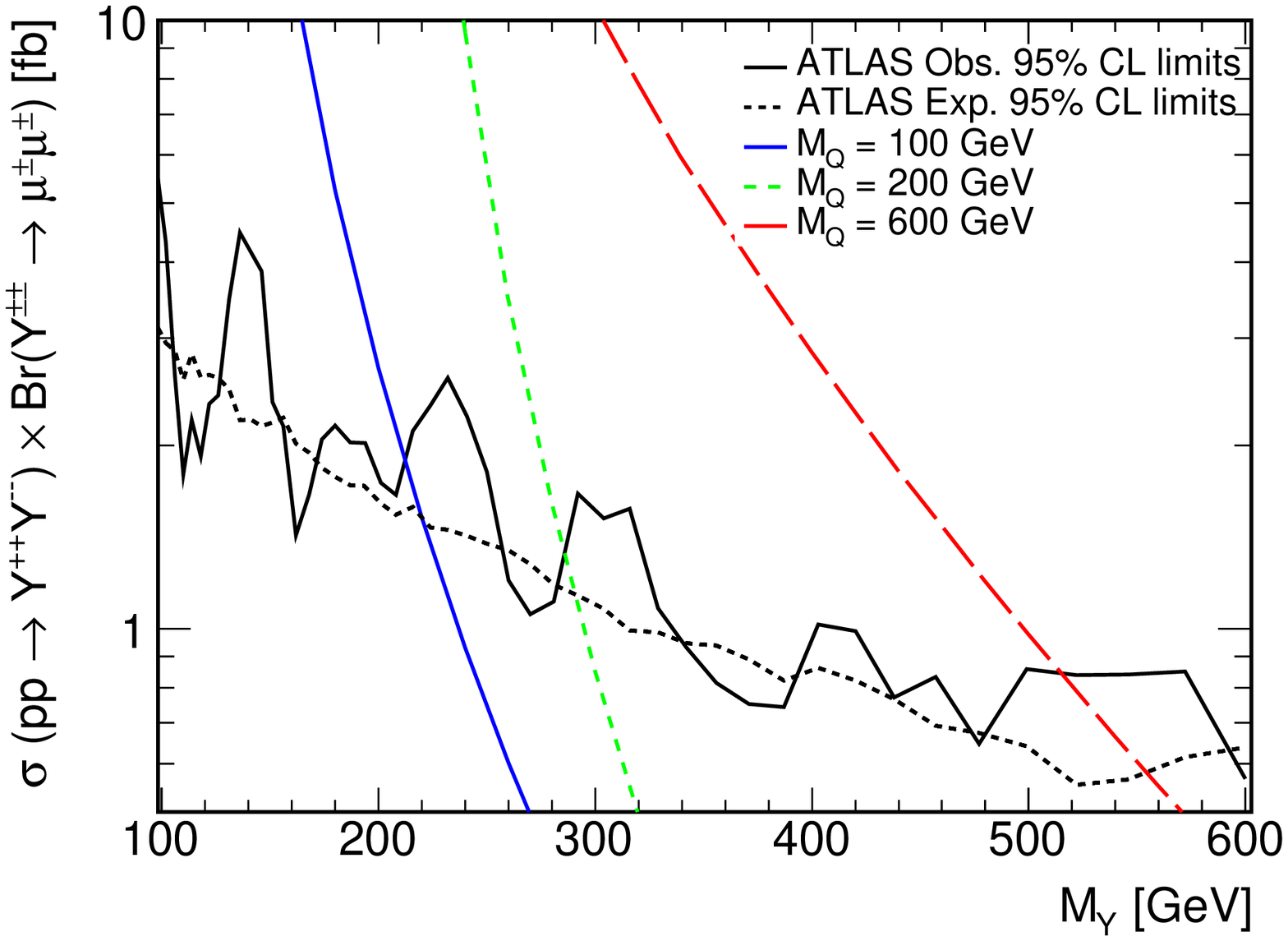}
  \caption{}
  \label{fig:sub1}
\end{subfigure}%
\begin{subfigure}{.5\textwidth}
  \centering
  \includegraphics[width=1.0\linewidth]{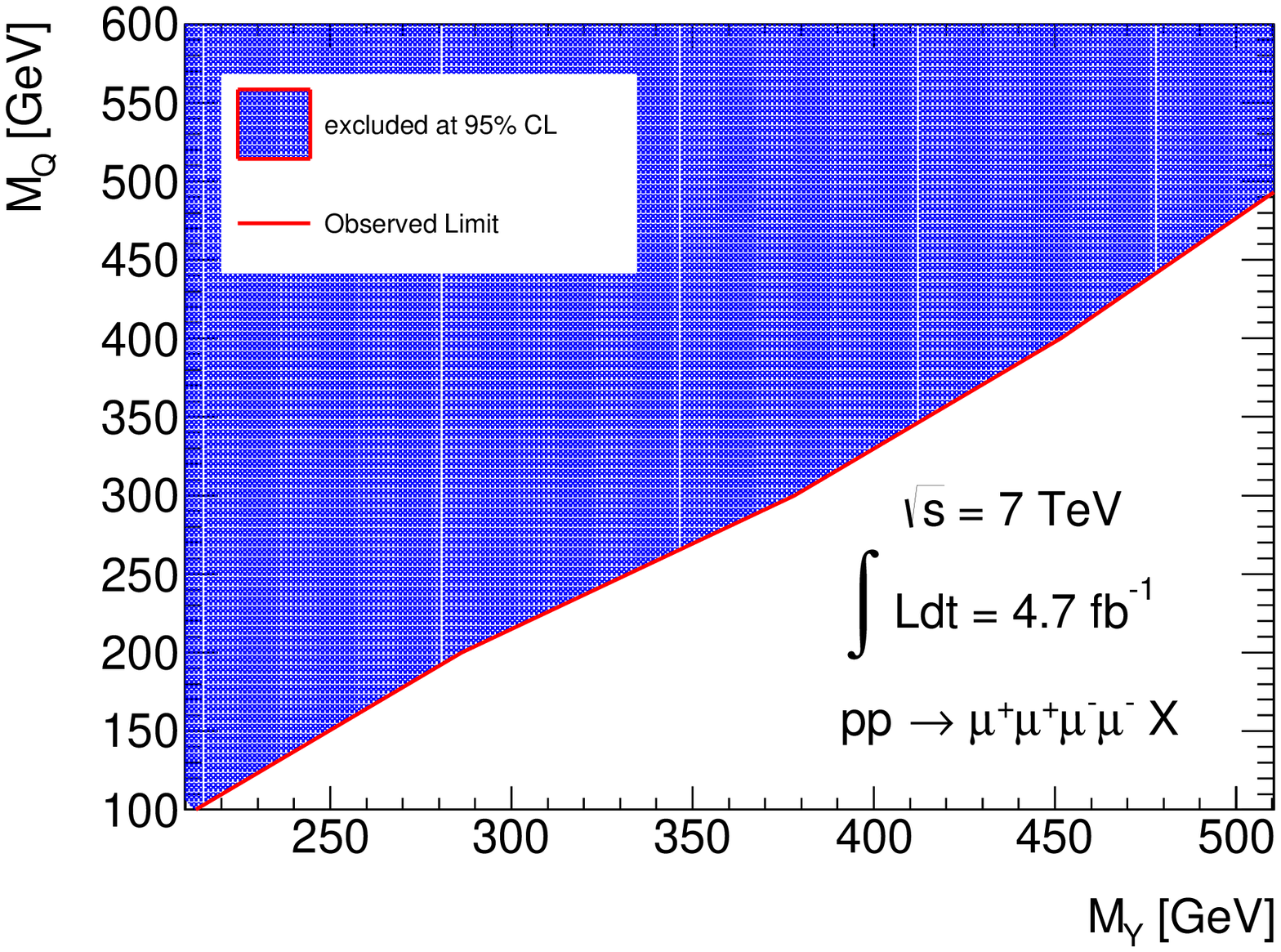}
  \caption{}
  \label{fig:sub2}
\end{subfigure}
\caption{(a)Upper limits on $\sigma \times Br$. 
  The black solid and dashed lines represent the ATLAS observed and expected limits, respectively. 
  The blue, green and red lines are the cross-section times branching ratio for bilepton production decaying into muons for 
  different values of the leptoquarks mass. (b) Exclusion region on the $M_Y \times M_Q$ plane.}
\label{fig:test1}
\end{figure}

The ATLAS collaboration have set upper limits on the cross-section for the doubly-charged Higss 
production in different final states at 7 TeV \cite{doubly_ATLAS}. 
The data sample correspond to an integrated luminosity of 4.7 $\textnormal{fb}^{-1}$.
The 95\% C.L. observed limits were placed as a function of the 
hypothesized boson mass, as shown in Figure \ref{fig:sub1}. These ATLAS limits were obtained for the number of 
leptons pairs originating from $H^{\pm\pm}$
in different mass windows. They  are converted to limits on the cross-section times branching ratio using the 
acceptance times efficiency derived from MC simulation. 
The ATLAS expected limits are the median values resulting from a large number of simulated 
pseudo-experiments assuming that no signal is present. Since bileptons are narrow resonances 
like the doubly-charged Higss (and therefore, they have similar acceptances and efficiencies), 
the ATLAS results can be used to derive limits on bileptons' masses.

\vskip .2 cm
The theoretical cross-section times  branching ratio for the process $p,p$ $\rightarrow$ $\mu^{+}\mu^{+}\mu^{-}\mu^{-} X$,
considering different bilepton and leptoquark masses, is calculated and compared with the cross-section limits obtained by ATLAS. 
The bilepton upper cross-section limit is derived from the intersection between the theoretical 
and the experimental curves. This limit is translated in the lower limit on the bilepton mass. 
Figure \ref{fig:sub1} illustrates the procedure for three different leptoquark masses. 
For $M_Q = 100$ GeV, vector bileptons with masses below 250 GeV are excluded. 
The strongest limit that can be derived for bileptons with 7 TeV data is $M_{Y} > 520$ GeV, 
corresponding to a leptoquark mass of $M_Q = 600$ GeV. Figure  \ref{fig:sub2} shows the exclusion region on the 
$M_Y \times M_Q$ plane,
obtained from Figure  \ref{fig:sub1}, considering six values of leptoquarks mass between 100 GeV and 600 GeV. 
The blue (dark) region is excluded at 95\%C.L.
These results agree very well with our prediction for 
bilepton exclusion with 5 $\textnormal{fb}^{-1}$ of data at 7 TeV (see Table II of Ref. \cite{Meirose&Nepomuceno}). 
Bileptons with masses between 250 GeV and 520 GeV,  depending on the leptoquark mass,  are excluded.


\section{LHC Run II: 13 TeV bilepton theoretical reach}
\label{sec:13TeV_discovery}

\begin{figure}
\centering
\begin{subfigure}{.5\textwidth}
  \centering
  \includegraphics[width=1.0\linewidth]{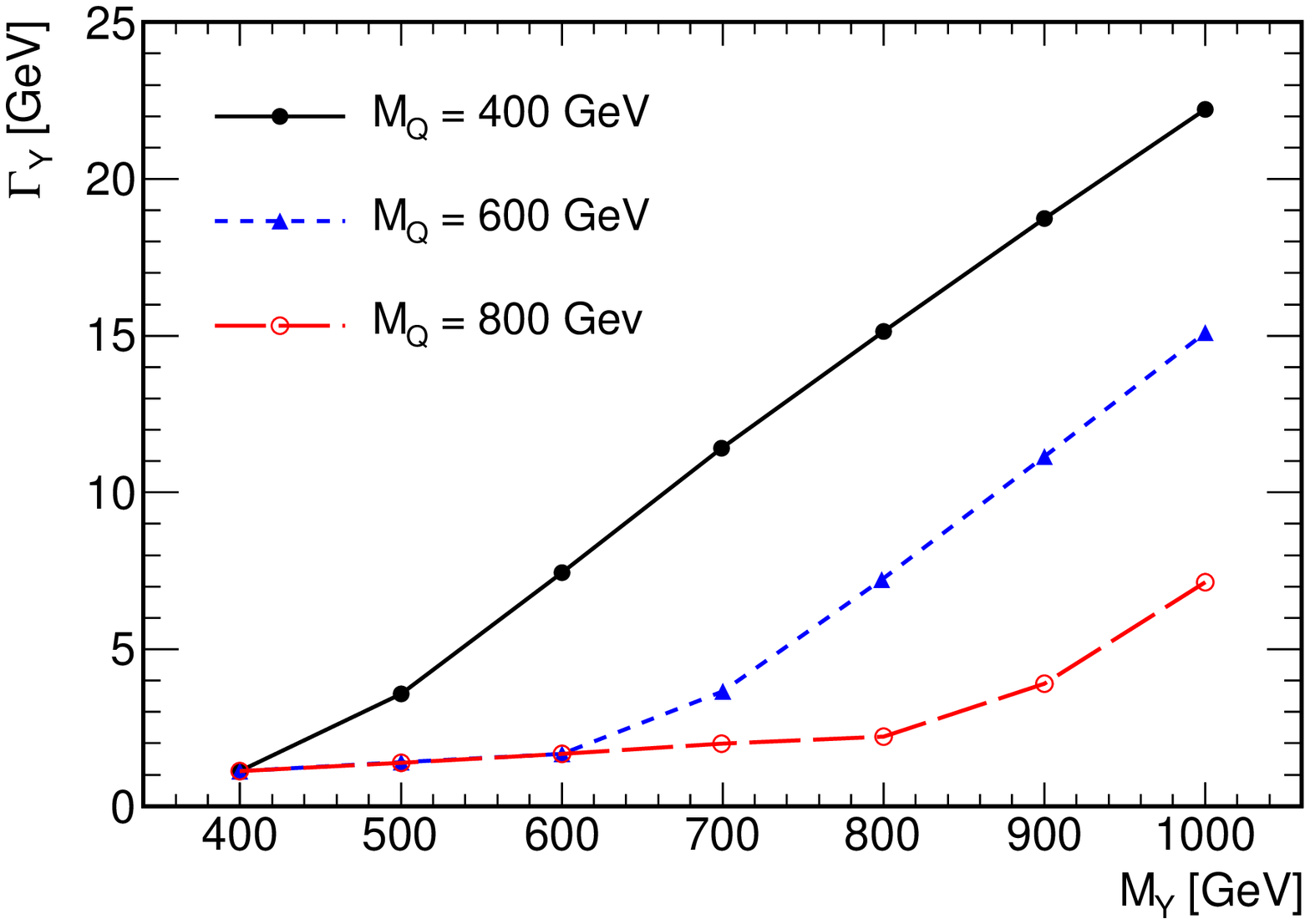}
\caption{}
  \label{fig:t2_sub1}
\end{subfigure}%
\begin{subfigure}{.5\textwidth}
  \centering
  \includegraphics[width=1.0\linewidth]{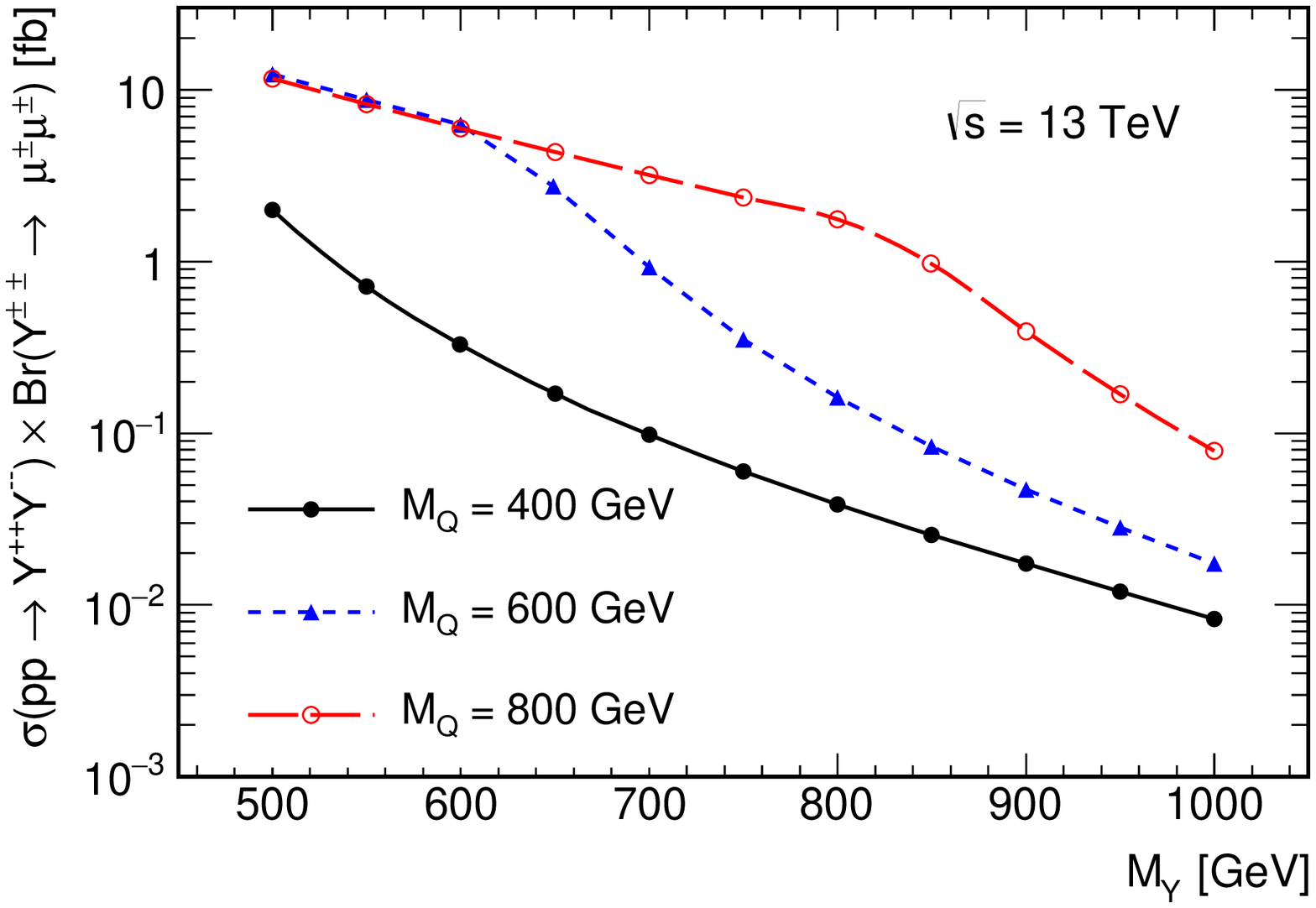}
  \caption{}
  \label{fig:t2_sub2}
\end{subfigure}
\caption{Bilepton width (a) and cross-section (b) for bilepton production at 13 TeV.}
\label{fig:test2}
\end{figure}

The LHC potential for discovering vector bileptons at a center-of-mass energy of 13 TeV is studied. 
Figure \ref{fig:test2} shows the bilepton width and cross-section for doubly charged bilepton production and subsequent decay to muons at 13 TeV 
for three different leptoquark masses. The values of bilepton and leptoquarks masses were chosen in a 
region beyond the region excluded.
Bileptons decays into leptoquarks explain the cross-section behavior observed for $M_Q = 600$ GeV and  $M_Q = 800$ GeV.
For $M_Y > M_Q$, bilepton decays like $Y^{\pm\pm} \longrightarrow qQ$
become kinematically allowed, which causes $Br(Y^{\pm\pm} \longrightarrow \ell^{\pm}\ell^{\pm})$ to decrease.

\vskip .2 cm
A fast detector simulation using \textsc{Delphes} is performed to estimate the acceptance and 
efficiency for reconstructing bileptons. In the analysis of the reconstructed events, at least four muons are initially selected. 
As each bilepton decays to a pair of same-sign muons, 
there are four muons in the final-state, two negatively and two positively charged. If more than four muons are found, 
the ones with higher transverse momentum ($\pt$) are chosen. 
All muons must be inside the detector acceptance ($|\eta|<$ 2.5) and have $\pt >$ 20 GeV. 
The  product of the acceptance and the selection efficiency after these cuts is around 80\%. 
As there is no trigger efficiency included in the simulation, we multiply the reconstruction 
efficiency by the expected trigger efficiency of 80\%  \cite{ATLAS_trigger}. The overall efficiency is then 64\%.

\begin{figure}
\centering
\begin{subfigure}{.5\textwidth}
  \centering
  \includegraphics[width=1.0\linewidth]{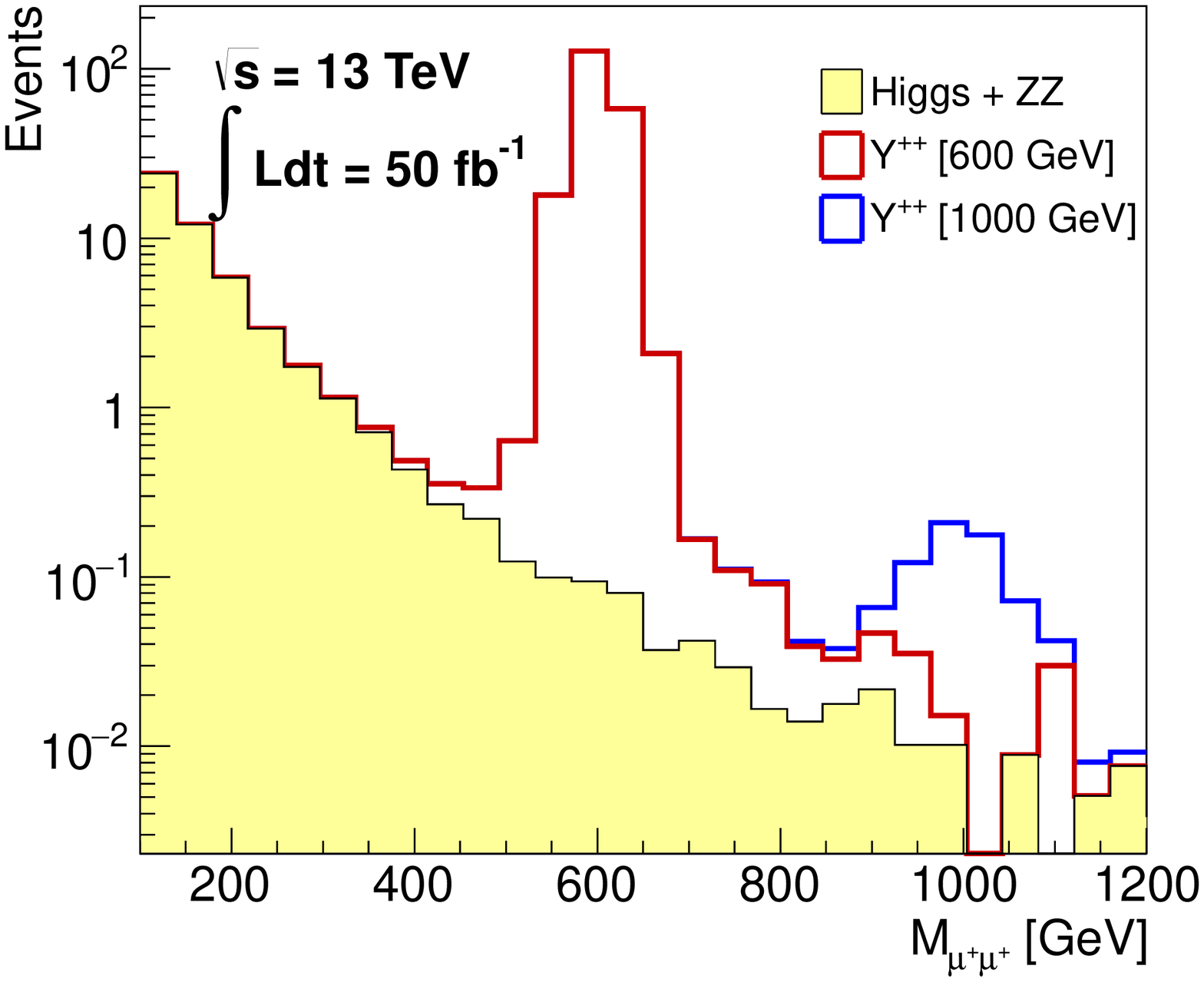}
  \caption{}
  \label{fig:t3_sub1}
\end{subfigure}%
\begin{subfigure}{.5\textwidth}
  \centering
  \includegraphics[width=1.0\linewidth]{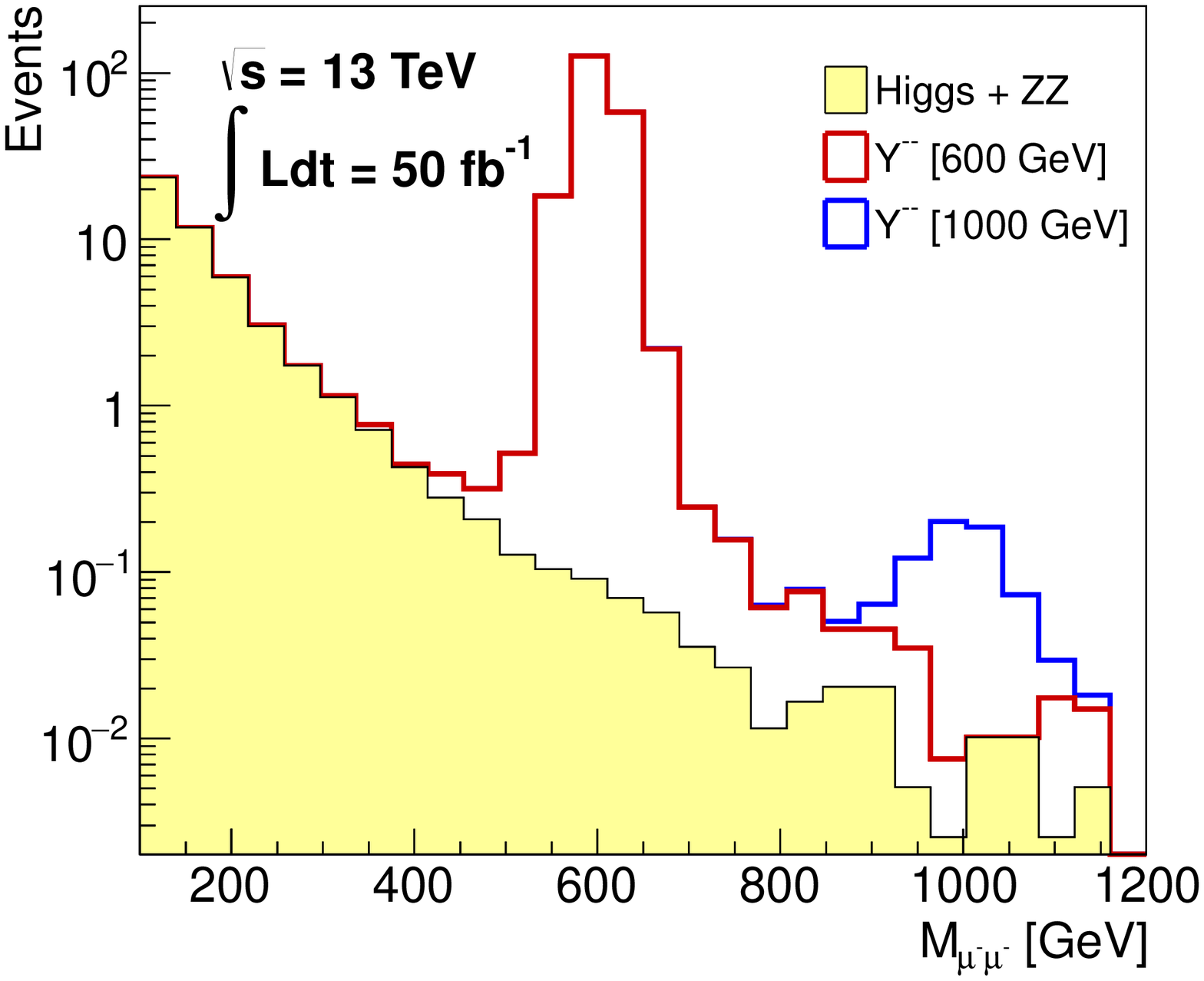}
  \caption{}
  \label{fig:t3_sub2}
\end{subfigure}
\caption{Invariant mass distributions for same-sign muons pairs produced by the background and by bileptons decay, assuming an 
integrated luminosity of 50 $\textnormal{fb}^{-1}$. The open red/blue histograms are two possibles positively charged (a) and 
negatively charged (b) bileptons signals.}
\label{fig:test3}
\end{figure}

\vskip .2 cm
The dominant background in this search are processes that can produce four muons in the final state. 
We have considered Higgs and $ZZ$ productions, both decaying to four muons. The Higgs background is found negligible above the muons invariant mass 
of 500 GeV. Figure \ref{fig:test3} shows the invariant mass distributions for each same-sign muons pair. 
The yellow histogram is the background, and the open histograms represent two possible bilepton signals. 
As the bileptons are produced in pairs, each same-sign muon pair have the same invariant mass distribution.  

\vskip .2 cm
The minimal integrated luminosity needed to discover a doubly-charged vector bilepton in the four-muon channel at LHC is calculated 
by comparing the background and signal invariant mass distributions through a chi-square analysis. 
The test is performed within a dimuon mass window of $[M_Y - 5\Gamma_{rec},M_Y + 5\Gamma_{rec}]$, where $\Gamma_{rec}$ is the width of 
a Gaussian fitted to the signal invariant mass distribution of the muons pairs. 
The bin width of the distribution is chosen so that it is larger than the invariant 
mass resolution determined from the detector simulation.  
For each value of bilepton and leptoquark mass, a hypothesis test is performed using as test statistic a chi-square given by \cite{Almeida}

\begin{equation}
 \chi^{2} = \sum_{i=1}^{n} \left [2(N_i - \nu_i) + 2(\nu_i+1)\textnormal{log}\left(\frac{2\nu_i+1}{2N_i+1}\right) \right]
\end{equation} 
\vskip .4 cm
where $n$ is the number of bins, $\nu_i$ is the background mean value, in the $i$th bin, determined 
from a large simulated sample, and $N_i$ is the number of events in each bin of the tested histogram. 
By conducting this analysis for 5000 MC pseudo-experiments, we can determine the chi-square distributions 
of the background-only and background plus signal hypotheses. 
The signficance level ($P$-value) is obtained by integrating the tail of the $\chi^2$ distribuiton
of the background-only hypothesis 

\begin{figure}
 \centering
 \includegraphics[width=.6\linewidth]{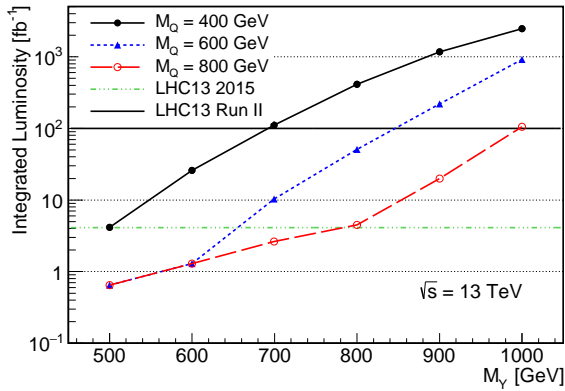}
 \caption{Minimal integrated luminosity needed for a 5$\sigma$ discovery of doubly-charged vector bileptons in the four muon final-state.}
\label{fig:test4}
\end{figure}

\vskip .4 cm
\begin{equation}
 P = \int_{\langle \chi^2_s \rangle}^{\infty} f(z) dz
\end{equation} 
\vskip .4 cm
where $\langle \chi^2_s \rangle$ is the $\chi^2$ most probable value for the background plus signal hypothesis. 
In order to estimate the amount of data needed to claim a bilepton discovery, 
the integrated luminosity is increased until we have $P < 3.0 \times 10^{-7}$, which corresponds to a significance of $5\sigma$. 
The results are shown in Figure \ref{fig:test4}. 

\vskip .2 cm
The horizontal dash-dot line in Figure \ref{fig:test4} represents the integrated 
luminosity delivered by the LHC in the first phase of the 13 TeV run ($\sim 4 \textnormal{fb}^{-1}$). 
Bileptons with masses up to 800 GeV can be probed with the available data. 
By the end of run II, with an integrated of 100 $\textnormal{fb}^{-1}$, 
bileptons with masses between 500 GeV and 1000 GeV could be discovered. 

\begin{figure}
\centering
\begin{subfigure}{.5\textwidth}
  \centering
  \includegraphics[width=1.0\linewidth]{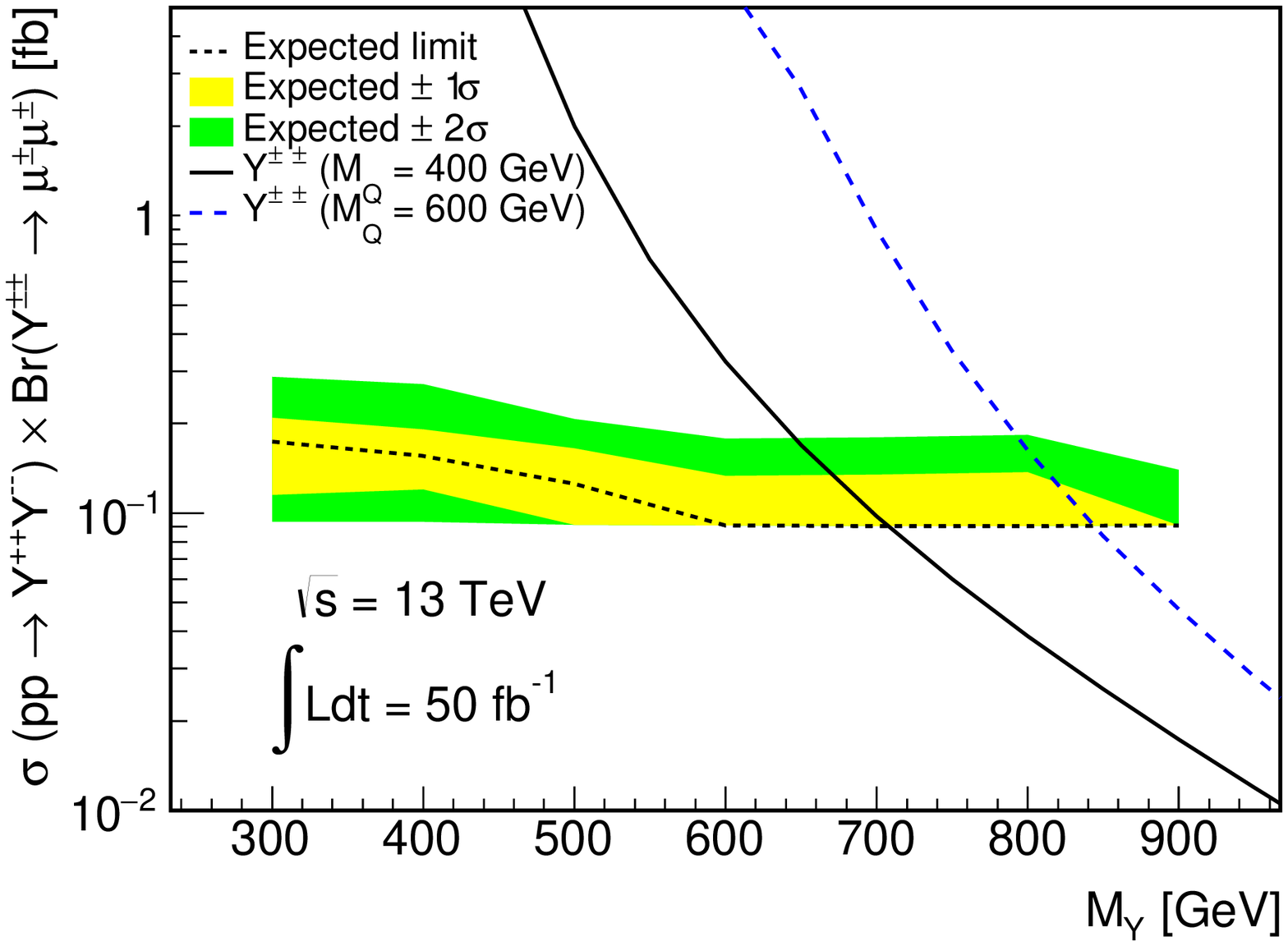}
  \caption{}
  \label{fig:t5_sub1}
\end{subfigure}%
\begin{subfigure}{.53\textwidth}
  \centering
  \includegraphics[width=1.0\linewidth]{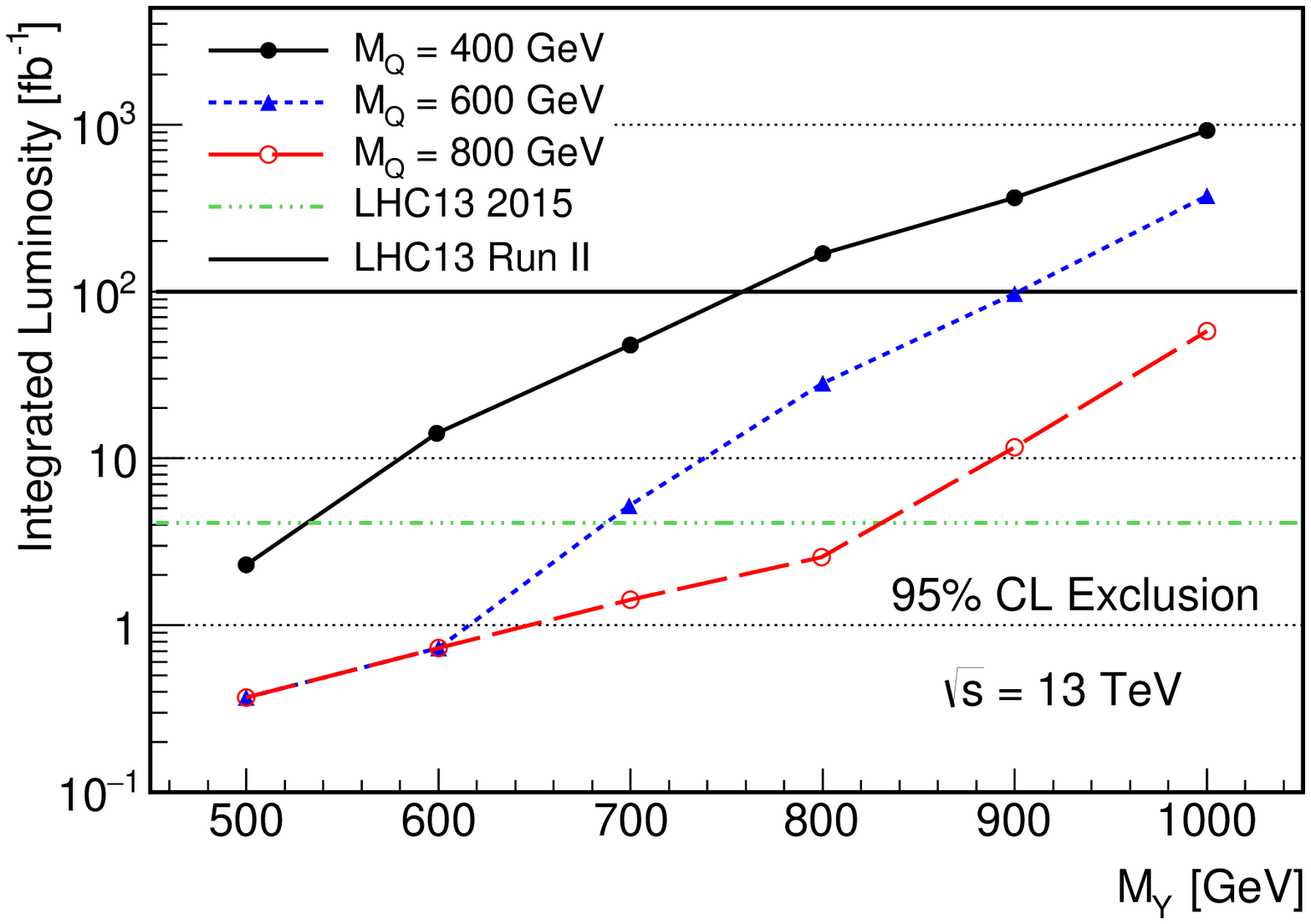}
  \caption{}
  \label{fig:t5_sub2}
\end{subfigure}
\caption{(a) Upper limits on $\sigma \times Br$ assuming 50 $\textnormal{fb}^{-1}$ of data at 13 TeV. 
(b) Minimal integrated luminosity needed to exclude bilepton of a given mass, for three different leptoquark masses.}
\label{fig:test5}
\end{figure}

\vskip .2 cm
If no bilepton signal is found at run II, the new LHC data can considerably extend the current limits of these particles. 
In order to calculate the exclusion limits that can be reached with a given integrated luminosity, a single bin analysis
applying a Bayesian technique is done.  An implementation of the method is available in the \textsc{Mclimit} program \cite{mclimit, Heinrich}. 
This approach assumes that the signal adds incoherently to the background. The inputs for the calculations are the expected 
number of signal and background events obtained from the detector simulation. 
Figure \ref{fig:t5_sub1} shows the expected limits on $\sigma \times Br$ assuming 50 $\textnormal{fb}^{-1}$ of data, for different bilepton mass hypothesis.  
The black dash line is the median values of the limits obtained from 1000 pseudo-experiments, and the yellow and green 
bands represent the $1\sigma$ and $2\sigma$ variation around the median, respectively. The lower bound $M_{Y} > 850$ GeV can be reach with this luminosity. 
This procedure is repeated for different values of the integrated luminosity, and for each of them, 
a lower mass limit for bileptons/leptoquarks is obtained. The results are shown in Figure \ref{fig:t5_sub2}. 
With 100 $\textnormal{fb}^{-1}$ of data, the bileptons limits can be extended above 1000 GeV in the 
most optimistic scenario. With 300 $\textnormal{fb}^{-1}$, the integrated luminosity expected for LHC's run III, 
masses up to 900 GeV can be excluded for the lowest branching ratio considered. 
In any case, one will still be below the theoretical upper limit $M_{Y} < $ 4 TeV imposed by the 331 model \cite{Frampton4TeV} . 



\section{Conclusions}
\label{sec:conclusion}

\vskip .2 cm
Exclusion limits on bileptons masses based on LHC real-data results at 7 TeV center-of-mass energy are derived.  
Bileptons masses in the range 250 $< M_Y <$ 520 GeV are excluded at 95\% C.L. The LHC potential to observe doubly-charged vector 
bileptons at 13 TeV center-of-mass energy in $pp$ collisions is also investigated. 
Taking into account reconstruction and trigger efficiencies of muon detection, bileptons masses between 500 GeV and 1000 GeV can
be observed by the end of run II. With the available data at 13 TeV, lower bounds from 530 GeV to 830 GeV can be estimated for the bilepton mass.
New data from run II can push the current limits up to 1040 GeV. 
Considering the theoretical constraint imposed by the 331 models on bilepton mass, 
our results show that the model cannot be fully excluded even at run III.

\vskip 1cm
{Acknowledgments} 
\newline
\noindent B. Meirose's work has been supported by US DOE Grant \\ 
Number DE-SC0010384. 
\newline
\noindent A. Nepomuceno thanks CNPq for the financial support.

\mbox{}
\nocite{*}
\bibliography{bibliography}
\bibliographystyle{JHEP}

\end{document}